# Optical studies of carrier and phonon dynamics in $Ga_{1-x}Mn_xAs$

K.J. Yee [a)] and D. Lee

*Department of Physics, Chungnam National University, Daejeon 305-764, Korea*

X. Liu, W. L. Lim, M. Dobrowolska and J.K. Furdyna

*Department of Physics, University of Notre Dame, Notre Dame, IN 46556*

Y.S. Lim

*Department of Applied Physics, Konkuk University, Chungju, Chungbuk 380-701, Korea*

K.G. Yee and D.S. Kim

*Department of Physics, Seoul National University, Seoul 151-747, Korea*

**Abstract**

We present a time-resolved optical study of the dynamics of carriers and phonons in $Ga_{1-x}Mn_xAs$ layers for a series of Mn and hole concentrations. While band filling is the dominant effect in transient optical absorption in low-temperature-grown (LT) GaAs, band gap renormalization effects become important with increasing Mn concentration in $Ga_{1-x}Mn_xAs$, as inferred from the sign of the absorption change. We also report direct observation on lattice vibrations in $Ga_{1-x}Mn_xAs$ layers via reflective electro-optic sampling technique. The data show increasingly fast dephasing of LO phonon oscillations for samples with increasing Mn and hole concentration, which can be understood in term of phonon scattering by the holes.

PACS : 75.50.Pp, 78.47.+p , 63.20.-e

The diluted magnetic semiconductor (DMS) $Ga_{1-x}Mn_xAs$ has attracted a great deal of attention because of the possibility which it offers of exploiting spin as an additional degree of freedom in electronic applications [1,2]. In $Ga_{1-x}Mn_xAs$ alloys, Mn atoms are incorporated at the Ga lattice sites as $Mn^{++}$ ions, acting as both magnetic moments and as acceptors. The ferromagnetic ordering of Mn spins, occurring at low temperatures, is described by the Zener model of ferromagnetism, in which the itinerant hole carriers mediate the ferromagnetic alignment of the magnetic ions [3]. With increasing Mn concentration $x$ in as-grown $Ga_{1-x}Mn_xAs$ samples, typically both the Curie temperature $T_C$ and the hole concentration $p$ have been shown to increase up to around $x = 0.06$, beyond which point $T_C$ and $p$ level off and decrease, primarily due to the formation of Mn interstitials [4-5]. In the Mn concentration range investigated in this paper ($0 < x < 0.06$), the total magnetic moment $M_S$ of $Ga_{1-x}Mn_xAs$, the Curie temperature $T_C$, and the hole concentration $p$ have all been shown to increase with $x$ [6].

Ferromagnetic $Ga_{1-x}Mn_xAs$ has already been investigated by a number of optical techniques, including optical absorption and magnetic circular dichroism (MCD); and Raman scattering was used to determine the hole concentration in this material [7-9]. However, in spite of their importance for basic understanding of the material as well as for spintronic device applications, studies of $Ga_{1-x}Mn_xAs$ by optical methods are much less fully explored than the investigation of this material by magnetic and transport techniques [4-5,10].

In this paper, we report a systematic optical study of carrier and phonon dynamics in $Ga_{1-x}Mn_xAs$ specimens as a function of Mn and hole concentrations. A series of $Ga_{1-x}Mn_xAs$/ZnSe heterostructures with different Mn concentrations $x$ were grown by molecular beam epitaxy (MBE) on semi-insulating (001) GaAs substrates in a Riber R&D MBE machine equipped with both III-V and II-VI growth chambers [11]. The Mn concentrations $x$ of the samples are 0.059 (sample A), 0.036 (B), 0.027 (C), and 0.008 (D), as deduced from the Mn cell temperature that has been previously calibrated by x-ray diffraction studies of $Ga_{1-x}Mn_xAs$/GaAs epilayers. The hole concentrations of the samples were estimated by measuring the Hall resistance at room temperature, where the contribution from the anomalous Hall effect, while not negligible, is considerably reduced. The estimated hole concentrations in samples A, B, C and D (in order of decreasing Mn concentrations) are $4.8 \times 10^{19}$, $2.0 \times 10^{19}$, $3.1 \times 10^{18}$ and $1.5 \times 10^{17}$ cm$^{-3}$, respectively. For $Ga_{1-x}Mn_xAs$ samples A and B, which have the highest Mn and hole concentrations, magneto-transport data indicate typical ferromagnetic behavior, with Curie temperatures of 54 K and 32 K, respectively. For sample D, which has a Mn concentration of less than 1%, no ferromagnetic behavior was observed down to 1.5 K.

To facilitate absorption measurements, the GaAs substrate was removed from the as-grown samples by mechanical polishing and chemical etching. Using a mode-locked Ti:Sapphire laser generating ultrashort pulses of 20 fs FWHM (full width at half maximum) at center wavelength of 780 nm (1.59 eV), we have performed transient absorption measurements at room temperature on the series of $Ga_{1-x}Mn_xAs$ layers prepared in this way, along with a reference low-temperature-grown (LT) GaAs sample. Electron and holes densities of $\sim 10^{17}$ cm$^{-3}$ were photo-excited by the pump pulse of 20 fs duration, and the carrier-induced absorption changes were measured by the weaker probe pulses as a function of time delay between the pump and the probe. To minimize coherent artifact signals that arise around zero time delay from direct nonlinear interactions between the pump and the probe photons, the probe polarization was oriented perpendicular to the pump polarization.

Figure 1 shows the time-resolved transient absorption changes (expressed in terms of the change $\Delta\alpha$ of the absorption coefficient relative to its value prior to excitation) for LT-GaAs and for the $Ga_{1-x}Mn_xAs$ samples. In the case of LT-GaAs, we see a sharp drop in absorption immediately following the excitation pulse, after which the absorption gradually recovers to its equilibrium (i.e., pre-excitation) value. The decrease of absorption through photoexcited carrier generation can be explained by the effect of band filling: i.e., the occupation of the conduction and valence bands generated by the pump pulse induces a decrease in the absorption probability of the probe beam that follows [12]. The nonequilibrium electron distribution created by the photo-excitation will eventually reach equilibrium (i.e., the state before pulsed excitation) through thermalization to the band edge states, as well as through interband recombination to the valence band. As we plot the LT-GaAs signal on logarithmic scale, two exponentially decaying components appear. The initial fast-decay term (with a 160 fs time scale) very likely arises from the relaxation process to the band edge; and the slower component (with a 1.6 ps time scale) can be ascribed to interband recombination and/or to electron trapping by defect states [13].

The transient absorption changes observed on $Ga_{1-x}Mn_xAs$ layers are distinctly different from those observed on LT-GaAs, and show a systematic dependence on the Mn concentrations. For $Ga_{1-x}Mn_xAs$ samples A and B (i.e., samples with a relatively high Mn content), the sign of absorption change is positive, i.e., absorption *increases* after pump excitation, opposite to the case of LT-GaAs. And rather than reflecting the non-equilibrium carrier density (which is maximum at the end of the pump pulse excitation), the absorption reaches a peak value at around 90 fs after the excitation, and then approaches equilibrium with a time scale of about 300 fs. This absorption increase due to carrier excitation must therefore represent a mechanism other than the band filling effect which characterized the LT-GaAs data. We suggest that the increase in absorption observed in $Ga_{1-x}Mn_xAs$ originates from bandgap renormalization, i.e., shrinkage of the bandgap due to correlated carrier effects which occurs as the electron density "spikes" at the pump pulse excitation, thus inducing an increase in absorption near the bandgap. This effect is generally much weaker than band filling effects in *ordered* materials [12], but may be comparable or even stronger when band filling is suppressed. In this connection we note that disorder – which tends to suppress band filling effects by relaxation of the k-selection rules – is naturally expected to increase in $Ga_{1-x}Mn_xAs$ with increasing *x*.

As we proceed to samples with a lower Mn concentration (samples C and D), the absorption first shows a sharp dip, followed by a clear increase over the equilibrium value. This mixed behavior characterized by dominant band filling immediately after the pump pulse, which is subsequently overwhelmed by the effects of bandgap renormalization on a longer time scale, is indicative that here we have a behavior with properties between those of the LT-GaAs and of $Ga_{1-x}Mn_xAs$ with high Mn content.

It is interesting that, while the band filling effect is maximum immediately following the pulsed excitation, band gap renormalization shows a delayed behavior. In Fig. 2 we show the time delay at which the absorption reaches its peak value in $Ga_{1-x}Mn_xAs$ layers. As the Mn concentration increases, the delay time at which the peak occurs systematically decreases, falling to about 90 fs for $x \approx 0.04$. This dependence is consistent with the thermalization time of hot electrons through scattering with an increasing concentration of holes. In this process, samples with higher hole concentrations are expected to have faster thermalization rates. From the fact that absorption change due to bandgap renormalization reaches a maximum after some time delay for thermalization, we infer that the contribution of band edge electrons to bandgap renormalization is greater than that of the optically activated hot electrons.

As we focus on the absorption changes at longer delay times, we notice a conspicuous modulation of the absorption signal. This modulation originates from coherent lattice vibrations. To gain better insight into this effect of lattice dynamics, we have performed reflective electro-optic sampling (REOS) experiments at room temperature on samples that did not undergo the etching processes. In REOS, coherent lattice oscillations excited by the pump beam are detected by measuring the rotation of the polarization of the probe beam caused by the collective lattice motions [14-15]. In Fig. 3a, we show the oscillating coherent phonon signals for each sample obtained by the REOS experiments. The oscillation period gives the phonon frequency, and the decay of the oscillation amplitude provides direct information on the scattering rate (dephasing) of the phonon mode. Regarding the dephasing rate, Fig. 3a shows that the samples with higher Mn content (A,B,C) have noticeably shorter dephasing times than those with lower Mn content (sample D), and than LT-GaAs.

We could extract the frequency and the dephasing time of $Ga_{1-x}Mn_xAs$ samples by fitting the phonon oscillations with an exponentially decaying sine function, $\Delta R/R = A*\exp(-t/\tau)\sin(\Omega t+\varphi_0)$, where A, $\tau$, $\Omega$, and $\varphi_0$ correspond to the amplitude, dephasing time, frequency, and initial phase, respectively. Figure 3(b) plots the longitudinal-optical (LO) phonon frequency and the dephasing rate ($1/\tau$) as a function of Mn concentration for the sample series used in this study. The red shift of the frequency with increasing Mn concentration is ascribed to the fact that the expansion of the lattice constant with increasing incorporation of Mn lowers the resonance frequency of the phonon modes. The observed behavior of the phonon decay rate, on the other hand, can be explained by dephasing of the phonons via the hole-phonon scattering processes: for samples having a higher Mn concentration and a higher hole density the phonon scattering by the holes will naturally become faster [16].

In conclusion, we have investigated carrier and phonon dynamics of $Ga_{1-x}Mn_xAs$ layers for a series of Mn and hole concentrations. We have shown that the transient optical absorption observed for LT-GaAs is determined by band filling effects, but in $Ga_{1-x}Mn_xAs$ the effects of bandgap renormalization become increasingly important as the Mn concentration increases, and eventually become dominant in samples with high Mn content. In samples with low Mn concentrations one observes the coexistence of both these processes, each distinguished by a different relaxation time. Additionally, coherent phonon oscillations were directly observed in the $Ga_{1-x}Mn_xAs$ samples by reflective electro-optic sampling experiments (REOS), showing a systematic progression of phonon dephasing with increasing Mn concentration. The hot electron thermalization effects observed in transient absorption as well as the dephasing of coherent LO phonons observed via REOS scale with the hole concentration in $Ga_{1-x}Mn_xAs$, leading us to suggest that both these effects result from scattering with the valence-band holes.

This work was supported by Korea Research Foundation Grant (KRF-2004-041-C00117). The Notre Dame part of this effort was supported by NSF Grants DMR02-

45227 and DMR-0210519.

a) Electronic mail: kyee@cnu.ac.kr

[1] H. Ohno, M. Munekata, T. Penney, S. von Molnar, and L. Chang, Phys. Rev. Lett. 68, 2664 (1992).

[2] H. Ohno, A. Shen, F. Matsukura, A. Oiwa, A. Endo, S. Katsumoto, and Y. Iye, Appl. Phys. Lett. 69, 363 (1996).

[3] T. Dietl, H. Ohno, F. Matsukura, J. Cibert, and D. Ferrand, Science 287, 1019 (2000).

[4] F. Matsukura, H. Ohno, A. Shen, and Y. Sugawara, Phys. Rev. B 57, R2037 (1998).

[5] H. Ohno, J. Magn. Magn. Mater. **200**, 110 (1999).

[6] We note for completeness that as $x$ continues to increase beyond the concentration investigated here, both $M_S$, $T_C$, and $p$ level off and eventually begin to drop due to formation of Mn interstitials [see K. M. Yu *et al.* in Phys. Rev. B **68**, 041308(R) (2003)].

[7] K. Ando, T. Hayashi, M. Tanaka, and A. Twardowski, J. Appl. Phys. 83, 6548 (1998).

[8] J. Szczytko, W. Bardyszewski, A.Twardowski, Phys. Rev. B 64, 75306 (2001).

[9] M. J. Seong, S. H. Chun, H. M. Cheong, N. Samarth, and A. Mascarenhas, Phys. Rev. B 66, 33202 (2002).

[10] J. K. Furdyna, X. Liu, W. L. Lim, Y. Sasaki, T. Wojtowicz, I. Kuryliszyn, S. Lee, K. M. Yu, and W. Walukiewicz, J. Korean Physical Society **42**, S579 (2003).

[11] X. Liu, Y. Sasaki, and J. K. Furdyna, Appl. Phys. Lett. 79, 2414 (2001).

[12] B.R. Bennet, R.A. Soref, and J.A. Del Alamo, IEEE J. Quantum Electron. 26, 113 (1990).

[13] A. Krotkus, K. Bertulis, L. Dapkus, U. Olin, and S. Marcinkevicius, Appl. Phys. Lett. 75, 3336 (1999).

[14] G. C. Cho, W. Kutt, and H. Kurz, Phys. Rev. Lett. **65**, 764 (1990).

[15] K. J. Yee, Y. S. Lim, K. G. Lee, E. Oh, D. S. Kim, and Y. S. Lim, Phys. Rev. Lett. **86,** 1630 (2001).

[16] Y.-M. Chang, Appl. Phys. Lett. **80**, 2487 (2002).

FIGURE CAPTIONS

Fig. 1. Time-resolved absorption changes $\Delta\alpha$ at room temperature for LT-GaAs and for $Ga_{1-x}Mn_xAs$ layers with different Mn concentrations. For clarity, the curves for successive values of $x$ are offset along the vertical axis, the pre-excitation level of each curve corresponding to the equilibrium absorption. The arrow on each curve indicates the position of maximum absorption. The inset shows the sequence of pump and probe pulses for a positive time delay.

Fig. 2. Time delay at which maximum absorption is observed in $Ga_{1-x}Mn_xAs$ as a function of Mn concentration x.

Fig. 3. (a) Coherent LO phonon oscillations at room temperature for LT-GaAs and $Ga_{1-x}Mn_xAs$ layers with different Mn concentrations. (b) Phonon frequency and the dephasing rate in $Ga_{1-x}Mn_xAs$ ($1/\tau$) as a function of Mn concentration $x$.

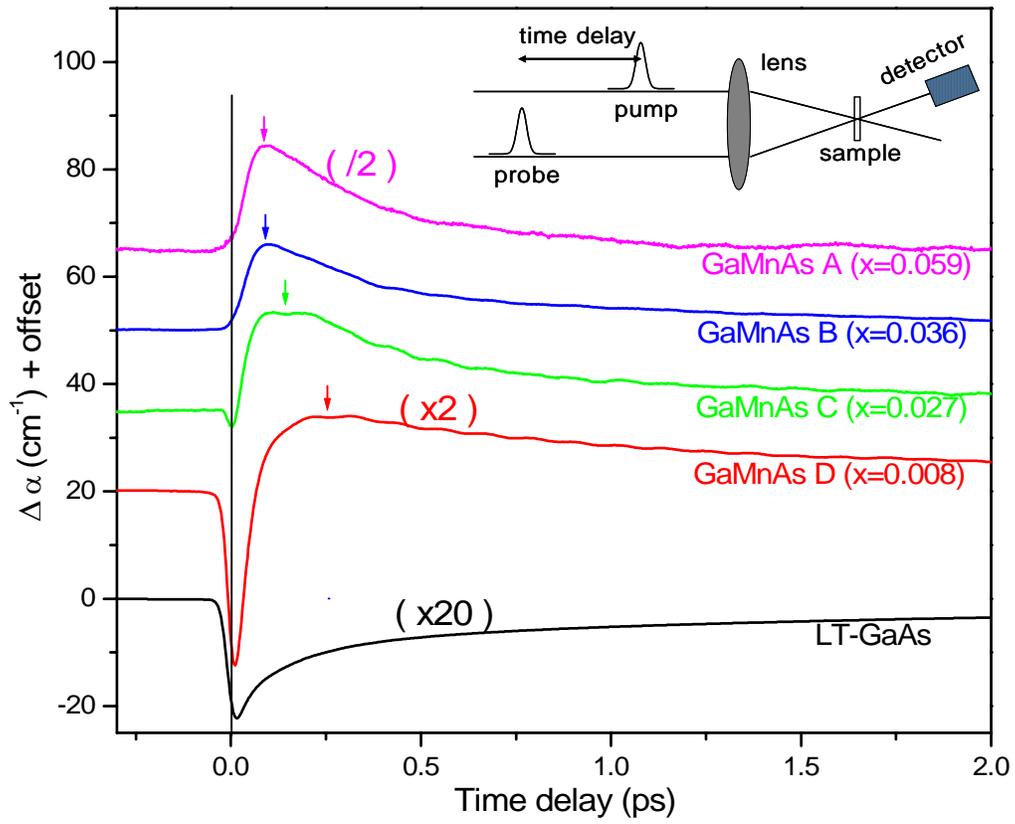

Figure 1

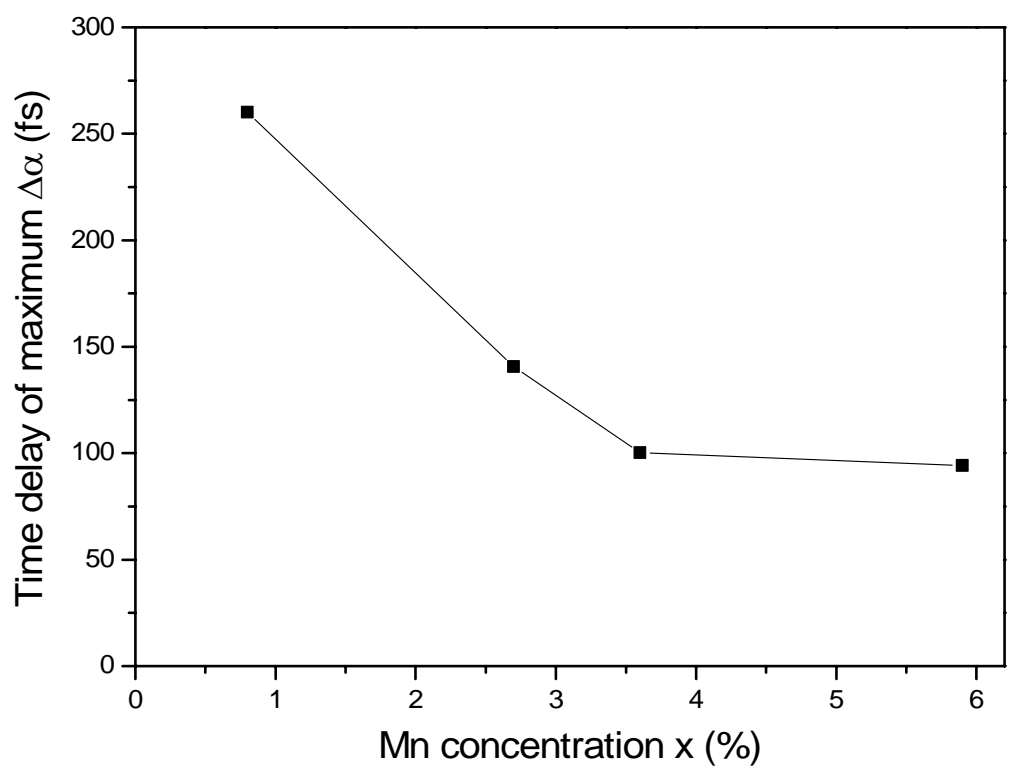

Figure 2

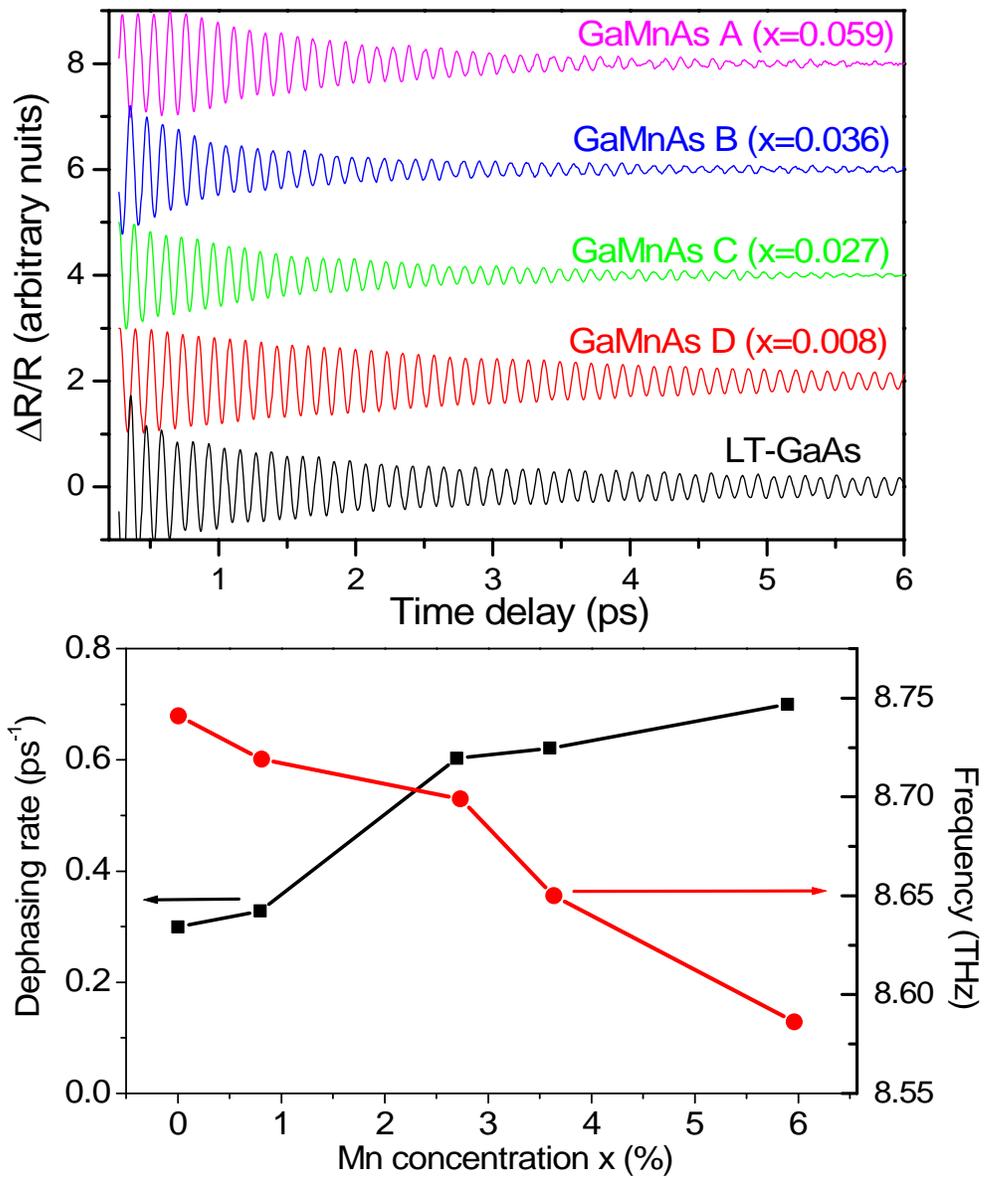

Figure 3